\documentclass[manuscript]{aastex}

\slugcomment{Submitted to Astrophysical J.}

\shorttitle{15 Febrary 2011  X2.2 Class Flare Observations}
\shortauthors{Gosain et al. 2011}

\begin{document}
\title{Evidence for collapsing fields in corona and photosphere during the 15 February 2011 X2.2 flare: SDO AIA and HMI Observations}
\author{S. Gosain\altaffilmark{1,2}
\altaffiltext{1}{National Solar Observatory, 950 N. Cherry Avenue, Tucson 85719, Arizona, USA.}
\altaffiltext{2}{Udaipur Solar Observatory, P. Box 198, Dewali, Udaipur 313001, Rajasthan, India}
\email{sgosain@nso.edu}}

\begin{abstract}
We use  high-resolution images of the sun obtained by the SDO/AIA instrument to study the evolution of the coronal loops in a flaring solar active region. During 15 February 2011 a X-2.2 class flare occurred in NOAA 11158, a $\beta\gamma\delta$ sunspot complex.  We identify three distinct phases of the coronal loop dynamics during this event: (i) {\it Slow rise phase}: slow rising motion of the loop-tops prior to the flare in response to slow rise of the underlying flux rope, (ii) {\it Collapse phase}:  sudden contraction of the loop-tops with lower loops collapsing earlier than the higher loops , and (iii) {\it Oscillation phase}: the loops exhibit global kink oscillations after the collapse phase at different periods, with period decreasing with decreasing height of the loops. The period of these loop oscillations is used to estimate the field strength in the coronal loops of different loop lengths in this active region. Further, we also use SDO/HMI observations to study the photospheric changes close to the polarity inversion line (PIL). The longitudinal magnetograms show step-wise permanent decrease in the magnetic flux after the flare over a coherent patch along the PIL. Further, we examine the HMI Stokes I,Q,U,V profiles over this patch and find that the Stokes-V signal systematically decreases while the Stokes-Q and U signal increases after the flare. These observations suggest that close to the PIL the  field configuration became more horizontal after the flare.  We also use  HMI vector magnetic field observations to quantify the changes in the field inclination angle and found an inward collapse of the field lines towards the polarity inversion line (PIL) by $\sim$ 10$^\circ$. These observations are consistent with the ``coronal implosion" scenario and its predictions about flare related photospheric field changes.

\end{abstract}

\section{Introduction}
The understanding of the coupling between the photosphere and solar corona requires simultaneous observations of the sun at different wavelengths. The observations from the recent Solar Dynamics Observatory (SDO) mission are best suited for such studies. The continuous high resolution observations of the full disk of the sun at a rapid cadence in multiple wavelengths allows us for the first time to study the evolution of solar active regions from photosphere to the corona on a regular basis. The evolution of photosphere as well as corona above the flaring active regions can give us important clues about the development of non-potentiality which fuels the flares. It was hitherto believed that flare was predominantly a coronal phenomena and that most of the flare related changes would be observed in corona with no visible changes at the photospheric boundary (Priest \& Forbes 2002). However, contemporary observations have shown that during large flares the abrupt changes are clearly visible at the photosphere (Kosovichev \& Zharkova 2001;Sudol \& Harvey 2005;Deng et al. 2005;Gosain et al. 2009b;Gosain \& Venkatakrishnan 2010;Petrie \& Sudol 2010).

Hudson et al. (2000) suggested that the build up of free energy in an active region  must lead to an inflation of the overlying coronal structure due to enhanced magnetic pressure. Further, they conjectured that the release of this free energy during flare and CMEs must consequently lead to a deflation of the magnetic field in the active region. This phenomena was termed as ``coronal implosion". Further, Hudson et al. (2008) predicted the consequences of the ``coronal implosion" at the photospheric boundary and suggested that the field inclination should change such that the final configuration is more horizontal. Recently, Wang \& Liu (2010) studied the vector field of 11 active regions during X-class flares and found that the post-flare  field configuration is  in agreement with the prediction of Hudson et al (2008). Further, similar pattern of magnetic field evolution was observed at the lower boundary in the 3D numerical MHD simulations of an erupting flux rope (Gibson \& Fan 2006; Fan 2010; Li et al. 2011). The observational signatures of coronal loop implosion or contraction were reported both during an erupting filament (Liu and Wang 2009a),  a C-class flare (Liu et al. 2009b) and a M-class flare (Liu \& Wang 2010a). However, these studies have been done in isolation and a combined study of flare related coronal as well as photospheric changes simultaneously during a flaring event has not been done due to lack of complete observations. Co-temporal HMI and AIA observations offers unique opportunity to detect such changes at an unprecedent temporal coverage.  Using recently released preliminary test vector magnetogram data for this active region Liu et al (2012) performed a nonlinear force free field (NLFFF) computation for this active region and found that mean horizontal field near flaring PIL increased by about 28 \% and the strong horizontal current system above this region collapsed downwards after the flare. Sun et al. (2012) also used these NLFFF computations and found that magnetic free energy reached about 2.6$\times 10^{32}$ ergs before the flare and about 0.3$\times 10^{32}$ ergs was released within 1 hour of X-class flare. Schrijver et al. (2011) used multi-wavelength coronal observations in AIA channels and studied the EUV wave during X-class flare in this region and found that the sections of  propagating coronal front running over quiet sun were consistent with adiabatic warming, while for other sections additional heating by Joule dissipation may be required.

Here we report on the observations of coronal implosion as observed in NOAA 11158 during X 2.2 class flare event during 15 February 2011. The detailed coverage of the event by AIA instrument onboard SDO allowed us to capture the evolution of coronal loops over this active region before, during and after the flare. We derive magnetic field strength estimates using the observed loop oscillations.  Also, the photospheric observations by HMI allowed us to study the evolution of photospheric line-of-sight (LOS) magnetic field component,  field inclination angle  as well as the direct observable i.e., Stokes profiles themselves in relation to this flare.  Further, we use SDO AIA images in EUV and detect a bright front emanating from this active region during flare. We report here a combined study of the coronal and photospheric changes during flare which are in agreement with the predictions by Hudson et al. (2000, 2008) and corroborate previous such observations carried out with limited spatial and/or temporal resolution.

In section 2 we describe the observational data and methods of analysis. Then in section 3 we present the evidence for coronal implosion as deduced from  the present analysis and the observed changes of the LOS magnetic field and the Stokes profiles at the photospheric boundary during the flare interval. We discuss these results in the light of previous reports as well as theoretical arguments.

\section{Observations and Data Analysis} 
\subsection{SDO/AIA Fe IX 171 \AA~ Observations} 
The first X-class flare of the current solar cycle (cycle 24) occurred during 01:48 UT on 15 February 2011 in a $\beta\gamma\delta$ sunspot complex, NOAA 11158. This event was observed in great detail by the instruments onboard  Solar Dynamics Observatory (SDO) (Pesnell et al. 2011). The photospheric observations were obtained by Helioseismic and Magnetic Imager (HMI) instrument (Schou et al. 2011) in Fe I 6173 \AA~ line (Norton et al. 2006) and the coronal observations were taken by the Atmospheric Imaging Assembly (AIA) instrument in as many as eight extreme ultraviolet (EUV) wavelength passbands (Lemen et al. 2011; Boerner et al. 2011). The coronal images analyzed here correspond to Fe IX 171 \AA~ passband. The level 1 SDO AIA images are obtained through the JSOC web server. The region-of-interest (ROI) used for the analysis is extracted from the level 1 full disk AIA images and  is shown in Figure 1, in inverted color map. The time interval chosen for the analysis is between 00:30 UT to 02:30 UT during 15 February 2011. The ROI is tracked in time using cross-correlation technique. Further, the intensity counts in the images are normalized for the variable exposure time (SDO AIA has variable exposure time during flare events). The loops (or bundle of loops) of interest are outlined by blue line segments and are marked 1 through 4. An artificial slit is placed along the yellow line such that the slit is normal to the loop apex. This facilitates the detection of any implosive motion in the loops. The space-time diagram corresponding to this slit is shown in the top panel of Figure 2, while the GOES X-ray flux in the two energy channels (1-8 \AA~ and 0.5 to 4 \AA~), during the same time interval, is shown in the bottom panel. The loops 1 through 4 are marked in the top panel of Figure 2 for identification. The three phases of loop dynamics discussed in next section are marked by orange colored arrows. The red dotted line marks the  slow-rise phase, the portion between two slanted yellow lines mark the implosion phase. The oscillation phase is visible towards the right portion of the space-time diagram. The dynamics of these loops is best viewed in the movie of this event which is provided as supplementary electronic material (movie1.mpg). The movie is created with by using Jhelioviewer software application (Mueller et al. 2010), provided by ESA/NASA.

\subsection{SDO/HMI Fe I 6173 \AA~ Observations} 
The photospheric observations by HMI onboard SDO are used for the present analysis. These observations are (i) longitudinal magnetograms, and (ii) Stokes profiles, S($\lambda$)=[I($\lambda$),Q($\lambda$),U($\lambda$),V($\lambda$)],  normalized to continuum intensity I$_c$. These are fulldisk  level-1 data obtained from JSOC web server. The HMI observations are taken in imaging spectro-polarimetric mode using Fe I 6173 \AA~ line. The Stokes profiles are obtained by tuning the filter across this line at six wavelength positions. The details of the HMI filter characteristics and calibration is described in Couvidat et al. (2011). The polarization calibration of the HMI Stokes level-1 data is described in Schou et al. (2010). The time interval of the observations used here  is same as for the SDO AIA images i.e., 00:30 to 02:30 UT. The cadence of longitudinal magnetograms is 45 seconds while that of Stokes profiles is 12 minutes.

 The longitudinal field corresponding to the ROI extracted from level-1 fulldisk LOS magnetogram data for our analysis is shown in Figure 3. Same ROI is extracted from the fulldisk Stokes images. For the registration of the longitudinal magnetograms and Stokes images (I,Q,U,V images) we perform cross-correlation of the HMI continuum intensity images (which are co-temporal with fulldisk LOS magnetograms) and Stokes-I (continuum) images and apply the derived shifts to these maps. This registration allows us to study the evolution of the Stokes profiles over the region of LOS field changes during the flare interval.  Further, we use the HMI vector magnetograms released by the HMI team (Hoeksema et al. 2012) for studying the changes in the field inclination angle near the flaring PIL. These vector magnetograms are available at a cadence of 12 minutes. We register the HMI vector magnetograms with the high cadence LOS magnetograms by cross-correlation method.

\section{Results} 
\subsection{The Evolution of the Coronal Loops} 
The four prominent loops (or bundle of loops) which show clear evidence for loop implosion (from the visual inspection of the movie)  are selected for analysis and are marked as 1,2,3,4 in the Figure 1. The apparent height of the loop apex ($H$) in the image plane is measured from the bright core of the active region  and is estimated to be $\approx$175, 155, 115, and 85 Mm for loops 1,2,3 and 4, respectively. The approximate loop length is then given by $L \approx\pi H$, as 550, 485, 360 and 267 Mm respectively for loops 1,2,3 and 4, assuming semicircular loop geometry. The space-time diagram corresponding to the artificial slit placed across the top (flat) part of four loops is shown in Figure 2. The dynamics of these loops shows three distinct phases of evolution. These phases are described below.

\subsubsection{Slow-rise phase:} During this phase we observe that the height of the apex of the loop 4 is increasing before the onset of flare. This increase in the loop height is perhaps present in all four loops but is distinctly visible for loop 4 owing to its higher intensity contrast.  The increase in apex height suggests that the loops are stretched vertically upwards due to increase in magnetic pressure in the core of the active region as a result of free energy buildup. We call it  slow-rise phase and it is marked by red dotted line in the space-time diagram in Figure 2. This  slow-rise  phase can be traced back upto 00:40 UT i.e., about one hour before the onset of flare. The slow-rise phase is gradual from 00:40 upto 01:25 UT and becomes rapid from 01:25 upto the onset of flare at $\sim$01:48 UT. The overall inflation of the loop height during the period 00:40 to 01:48 UT is about 10 arc-sec which corresponds to $\sim$7 Mm.

\subsubsection{Collapse phase:} During this phase all the four loops show a sudden implosive decrease in their apex height. Such visible change in apex height of loops could also be attributed  to the change of the tilt angle of the loop's plane to the line-of-sight, due to flare impulse. However, in that case we would expect the loops to restore back their apex heights once the flare impulse has passed the loops. In the space-time diagram (Figure 2.) however, we can notice that the change in apex height is permanent and the loops oscillate (oscillation phase described below) about the new mean apex height (blue arrows) of the loops which is lower than the pre-flare apex height of the loops (i.e., the height at the beginning of implosion phase, first yellow line) . Thus we can safely believe the loop dynamics to be polarized essentially in the loop's plane. The observed implosion phase is marked by the two yellow lines. The first yellow line joins the epochs when first signature of implosion can be traced for the loops 1-to-4, while the second yellow line joins the epochs when the maximum contraction for the loops 1-to-4 is reached. It may be noticed that there is a time delay between the onset of implosion phase of the loops, with lower loops imploding earlier than the higher ones. This is discussed further in section 4 where we present the possible scenario for the observed delay in the contraction of the loops.

 Moreover, it may be noticed that the loops 1-4 continue to rise for few minutes after the flare onset. A plausible explanation for this observations may be as follows.  The flare onset suggests  the onset of magnetic reconnection beneath the rising flux rope, while  the continued expansion of the observed overlying loops for minutes after the flare onset suggests that the part of the flux rope underneath these loops 1-4  is still expanding. It may be noted that the foot point of an eruptive flux rope remain anchored in the photosphere. Since the observed loops 1-4 are not exactly over the central portion of eruptive flux rope but over the peripheral part of flux rope, they probably continue to rise for some time as a part of the evolution of the foot-points of an eruptive CME flux rope.

\subsubsection{Oscillation phase:} After the implosion phase the loops begin to oscillate about their new contracted position, which we call as the oscillation phase. It may be noticed that the periods of the oscillations are not the same for the loops 1-to-4. By inspection of space-time diagram we deduce the oscillation period in the range of 660 seconds for loop 1 and 2, and  about 450  and 180 seconds for loops 3 and 4, respectively. Since these oscillations show transverse displacement of the entire loop with respect to loop position we believe these are the fundamental (first harmonic) fast kink mode MHD oscillations. The phase speed of this mode,$C_K$, in the low plasma $\beta$ limit and the assumption that the loop width is much smaller than loop length, is given as (Roberts et al. 1983) $$C_K\approx(\frac{2}{1+\rho_e/\rho_i})^{1/2}C_A$$
where $C_A$ is the  Alfv\'{e}n speed in the loop and $\rho_e/\rho_i$ is the ratio of densities outside and inside of the loop. The value of $C_K\approx 2L/P$ for the loops 1-4 can be determined from the observations as 1670, 1470, 1600 and 2966 km/s. The parameter $\rho_e/\rho_i$ is unknown and can take typically values from 0 to 0.3. Following, Nakariakov et al. (1999) we take $\rho_e/\rho_i$ to be 0.1 and compute the Alfv\'{e}n speed using the relation above as 1238, 1090, 1186 and 2199 km/s for loops 1-4 respectively. Then using relation $V_A=2.18\times10^{11} B/\sqrt{n}$ we estimate the field strength inside the loops 1-4 to be in the range 15-43, 13-38, 15-41, and 28-77 G for loops 1-4, respectively.   Where, the upper and lower limits corresponds to upper and lower limit of the estimated loop density $n$$\simeq$ $10^{9.33\pm0.44}$ cm$^{-3}$, by Aschwanden et al. (2011) for a sample of 570 loop segments in this active region using automated DEM (Differential Emission Measure) analysis of AIA observations. These values of the magnetic field strength are consistent (by the order of magnitude) with the previous seismological estimations using TRACE observations (Nakariakov \& Ofman 2001). It may be noted that the field strength in loop 4 which is lower in height is typically higher than the field strength derived for higher loops 1-3, which is expected as the magnetic field strength decreases with height due to expansion of the magnetic field in the coronal volume above the photosphere. A more detailed investigation of these oscillation along with their damping times is deferred to another paper, where we plan to compare the field strengths deduced here with the results of  force-free field (FFF) extrapolation.

\subsection{The Evolution of the Photospheric Magnetic Field} 
In this section we analyze the evolution of photospheric magnetic field for this active region using the longitudinal magnetograms obtained from SDO HMI. In addition to the evolution of the LOS field  we also show the evolution of the Stokes I,Q,U,V profiles themselves at the locations where we detect longitudinal field changes, together with the evolution in a region away from the flaring region, which serves as a control region for comparison.

\subsection{Evolution of Longitudinal Field during Flare} 
The high-resolution  longitudinal magnetograms available from HMI at a cadence of 45 seconds are used to study the evolution of the field during the flare interval.
 The magnetograms of the field-of-view is shown in the top panel of Figure 3.  The changes in the absolute value of LOS magnetic flux is obtained by taking the  running difference of the magnetograms (absolute value) observed between 01:45 and 02:00 UT. The difference image  is shown in the bottom panel of Figure 3. The positive changes (white areas) correspond to the decrease in LOS flux while negative (black areas) correspond to the increase in LOS flux. It may be noticed that in the periphery of active region the random black and white differences dominate, which are due to the small scale evolution and convective drift. While near the PIL there is a coherent patch showing a decrease of the LOS flux. Overall it may be noticed the areas of LOS flux decrease dominates over LOS flux increase.  Further, we study the nature of evolution of the LOS magnetic flux in two regions, one located near PIL and one away from the flaring region, these regions are marked by box 1 and 2 in Figure 3. The time profile of the evolution of LOS flux in the two boxes is shown in Figure 4. The bottom panel shows the evolution of LOS flux in box 1 (quiet region) and top panel shows the evolution in box 2 (flaring region). An abrupt change in the LOS flux can be noticed near the flaring region which is permanent and is distinct from the gradual evolution of LOS flux in the control region (box 1).   The earlier studies of LOS flux changes during strong X-class flares using low resolution observations obtained from ground based GONG and space based SOHO MDI instruments have shown such behavior of  LOS flux during strong flares (Kosovichev \& Zharkova 2001; Sudol \& Harvey 2005; Petrie \& Sudol 2010). The dominance of the regions of decreasing flux over increasing flux is consistent with earlier results (Petrie \& Sudol 2010).

\subsection{Evolution of Stokes profiles during Flare} 
For the 15 February 2011 X2.2 flare there are other photospheric changes that have been reported, for example, (i)  Kosovichev (2011) reported the detection of a powerful ``sunquake", (ii) Maurya et al (2011) reported detection of transient doppler and magnetic signatures associated with flare ribbons and (iii) Wang et al. (2011) reported a permanent enhancement of the linear polarization signals near the PIL which was interpreted as an enhancement of transverse magnetic field. To add to these photospheric observations we study the evolution the spatially averaged full Stokes profiles observed by HMI instrument over the rectangular boxes (2.5\H{}$\times$2.5\H{}), labeled 1 and 2, displayed in the panels of Figure 3. The box 1, is located in a quiet magnetic region away from the flaring region, while the box 2, is located over the coherent patch where the LOS flux decreased during the flare interval and is close to the PIL.

Figure 5 shows the Stokes I,Q,U and V profiles, normalized to continuum intensity, during the pre-flare times 01:36 and 01:48 UT (black and blue curves) and post-flare times 02:00 and 02:12 UT (green and red curves). It can be clearly noticed that the post-flare Stokes-Q and U profiles show an enhancement in the signal while the Stokes-V profile shows a decrease in the signal. This behavior of the Stokes profiles clearly indicates that the post-flare field changes orientation from vertical to more horizontal. This behavior is consistent with the interpretation by Wang et al. (2011) that the field near PIL becomes more horizontal after the flare, which was based on their study of the trend in linear polarization signal alone. It may be noted that an enhancement of linear polarization alone is not sufficient to conclude the orientation of field becomes more horizontal, we need to monitor all Stokes parameters simultaneously specially the relative variation of linear and circular polarization components. As seen in Figure 5, the relative variation suggests beyond doubt that the field inclination has changed from more vertical to more horizontal.  In the next section we use the preliminary vector magnetograms released by the HMI team  (Hoeksema et al. 2012) to quantify the amount of change in the inclination angle over this patch during the flare interval.

\subsection{Evolution of Field Inclination during Flare} 
Recent study of the HMI vector field and its non-linear force-free field (NLFFF) extrapolation reveals an increase in the horizontal field component by about $\sim$ 28\% near the PIL (Liu et al. 2012). While horizontal field component comprises of field strength and inclination, ($B_h=B sin\gamma$), here we examine how does the field inclination which is indicative of the field topology, change during the flare interval across the PIL. Using these data sets in a statistical sense, Sun et al (2012) show that in a box near PIL the distribution of inclination angles shifts towards more inclined fields. Similarly, Wang et al. (2012) studied temporal evolution of inclination near PIL and found a permanent change in the inclination of the fields towards horizontal. In Figure 6 we show maps of magnetic field inclination angle from HMI vector field data and their running difference corresponding to times 01:48 and 02:00 UT. A coherent patch of changing inclination can be seen near the PIL. The patch has opposite sign on either side of the PIL which means that the field on either side of the PIL becomes more horizontal since the inclination angle is measured in 0-180$^\circ$ range, where 90$^\circ$ corresponds to horizontal field and 0$^\circ$ and 180$^\circ$ correspond to field pointing upward (positive polarity) and downward (negative polarity), respectively. We select a rectangular box over this region where we see coherent change in inclination angle and plot its profile averaged in the x-direction so that a mean variation across the PIL can be estimated. The profile is shown in the right panel of Figure 6. It may be noticed that the change of $\sim$10$^\circ$ towards horizontal direction is seen on either side of the PIL, suggesting an inward collapse (towards PIL) of the field configuration. This is consistent with the predictions about the possible changes in field topology after a major flare by Hudson et al. (2008).

\section{Discussion and Conclusions}
The observations from SDO provides the best spatial and temporal coverage to study the evolution of flaring active regions. In the present work we studied the evolution of coronal loops as well as the photospheric magnetic field and Stokes profiles in relation to a X2.2 class flare.  The  flare related changes in the active region corona and the photospheric field are found to be in agreement with the predictions made by Hudson et al. (2000, 2008) using free energy arguments. The two main predictions namely the implosion in the corona and the changes in the magnetic field configuration at the photospheric boundary are tied to the argument that the post-flare state must correspond to a lower energy state as compared to the pre-flare state, irrespective of the detailed mechanism by which the excess energy is released. These predictions have been tested  in the past using either coronal or photospheric observations separately. In the present case we were able to observe both of these predictions  in a single active region during a X-class flare, thanks to excellent spatial, temporal and spectral coverage by the SDO mission.

 The three phases of the evolution of the coronal loops above the active region during the flare interval could be resolved in great detail in the present case. The slow-rise phase during which the loop height is observed to increase prior to the flare/CME onset has been studied earlier by Liu et al. (2010c) in an another event. Such an expansion could happen due to the slow rising of a flux rope  as seen in Liu et al. (2010b), which is often invisible unless being traced by filament material as in Gosain et al. (2009a). Schrijver et al. (2011) studied this active region in detail and identified a erupting flux rope structure during the flare. They detected a EUV wave associated with the eruption. The expanding overlying loop system in response to rising and expanding flux rope in the core of eruptive active regions is a common phenomena and in the present case with high spatial and temporal resolution we can observe the phenomena in great detail.  A plausible explanation for the observation of successive collapse of loops with increasing height in Figure 2 can be as follows. According to the model of flare given by Hirayama (1974) (see their Figure 1 (b)), reconnection sets-in in the current sheet formed below the rising flux rope and the compression region is formed causing the surrounding field to collapse inwards. As the flux rope moves forwards the centroid of the compression region will also move upwards and so one would expect collapse of surrounding fields successively at greater heights. This may explain our observation of a systematic delay in the onset of collapse of the loops with height. The EUV bright fronts are a signature of expanding flux rope (Schrijver et al 2011). In Figure 7 we show a running difference of EUV AIA 171 \AA\ images during flare interval. In Figure 7(a) we mark the EUV bright front observed in its early phase. During this time the implosion in loops (1-4) has not started. Further, the Figure 7(b) shows the location of EUV bright front at an instant when the first signature of collapse of loops (lower loop system, 4 in Figure 1) can be traced in the running difference. The collapse of higher loops (loops 1-2) appears later in time, as shown in  Figure 7(c), when the flux rope rises to greater height. Such a scenario should be tested with more such observations of powerful flares as well as with numerical MHD simulations of solar eruptions.


The slow-rise phase of overlying active region loops which lasts several minutes before the flare occurrence could be a precursor to major flares and therefore has a potential for flare detection. The realtime image processing and analysis tools could be used for detecting such precursors in flares. The implosive phase is clearly visible in a hierarchy of loops during the flare interval. Similar observations of implosive phenomena for other flares observed by SDO AIA would be useful to establish the phenomena further. The spatial variation of implosion over active region could be analyzed in detail to investigate the locations of energy storage in active regions. Further, the oscillation phase seen in loops analyzed here as well as other loops which are not analyzed here but can clearly be seen in the movie (online material) can be exploited in detail by the methods of coronal seismology to probe the properties of coronal loops (for example magnetic field) over active regions. We plan to carry out a detailed study of the oscillation period and damping times of various loops for this active region in a future work.

Further, the evolution of LOS magnetic field  as well as the observed behavior of the Stokes profiles themselves firmly establish the prediction (Hudson et al. 2008) that the field lines in the aftermath of flare become more horizontal. Such changes of photospheric field i.e., field becoming more horizontal after the flare were also found to be statistically significant in a recent study of 11 X-class flares by Wang \& Liu (2010), where they found in most of the cases transverse field near the PIL becoming more strong.

In conclusion we could study the signatures of imploding field configuration in coronal images as well as photospheric magnetic field. More such observations during powerful flare would be helpful in establishing the relationship between various flare related parameters like the relation between the extent of loop contraction, change in field topology near the PIL, peak X-ray output of the flare, the kinetic energy of the accompanied CMEs and the helioseismic response of a flare. Such comprehensive studies have not been undertaken in the past due to limited coverage of flaring events but with SDO , STEREO and Hinode  observatories there is a good scope for performing these studies which would eventually help in testing theoretical ideas about the dynamics of the solar flares (Hudson et al. 2011).

\acknowledgments
 I thank the anonymous referee for useful comments and suggestions that improved the presentation of this work. Also, I thank Dr. Valery Nakariakov and Dr. Pascal D\'{e}moulin for fruitful discussions at The Sun 360$^\circ$ Meeting at Kiel, Germany. The data provided under open data policy of Solar Dynamics Observatory (SDO) mission is acknowledged. Also, I thank the efforts of HMI and AIA teams, supported by NASA and LMSAL, for developing the excellent instruments. The work was carried out at National Solar Observatory, Tucson, Arizona, USA  while the author SG was on Extraordinary Study Leave from Udaipur Solar Observatory, Physical Research Laboratory, Udaipur, India.

\clearpage

\begin{figure}
\epsscale{0.70}
\plotone{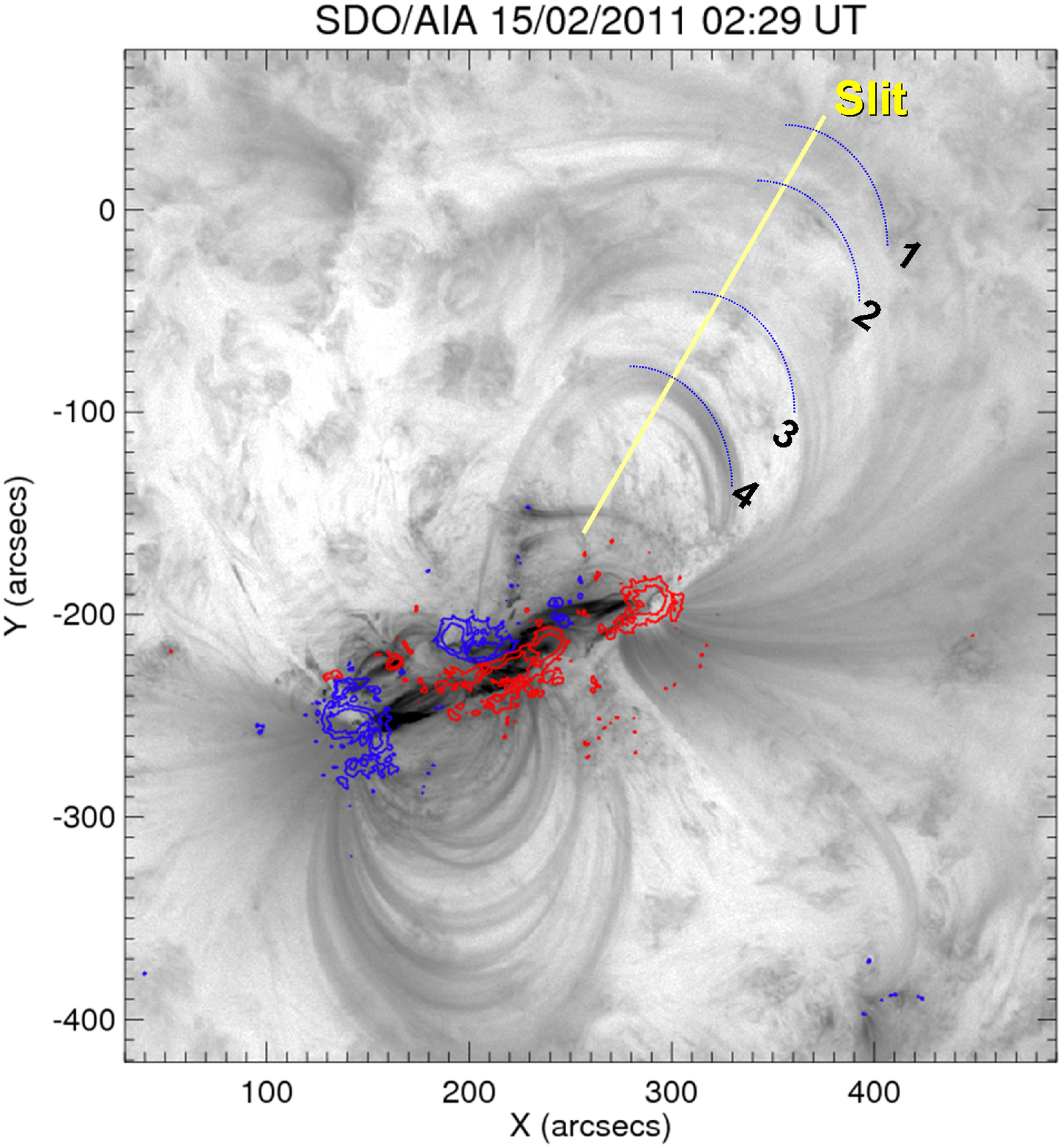}
\caption{The inverted color map of the active region NOAA 11158 observed in Fe IX 171 \AA~ wavelength by SDO AIA instrument during 02:29 UT on 15 February 2011. The loops marked 1-4 are studied for temporal evolution and are highlighted by blue curved line segments.  The line contours at 500 and 1000 G level of the longitudinal magnetic field observed by SDO/HMI instrument are overlaid in blue (red) colors, representing negative (positive) polarity, respectively.  The yellow line marks the position of the artificial slit that is placed to sample the dynamics of the apex of the loops. The  space-time diagram corresponding to the slit is shown in Figure 2. \label{fig1}}
\end{figure}

\clearpage

\begin{figure}
\epsscale{1.0}
~~~~~~~~~~~~\plotone{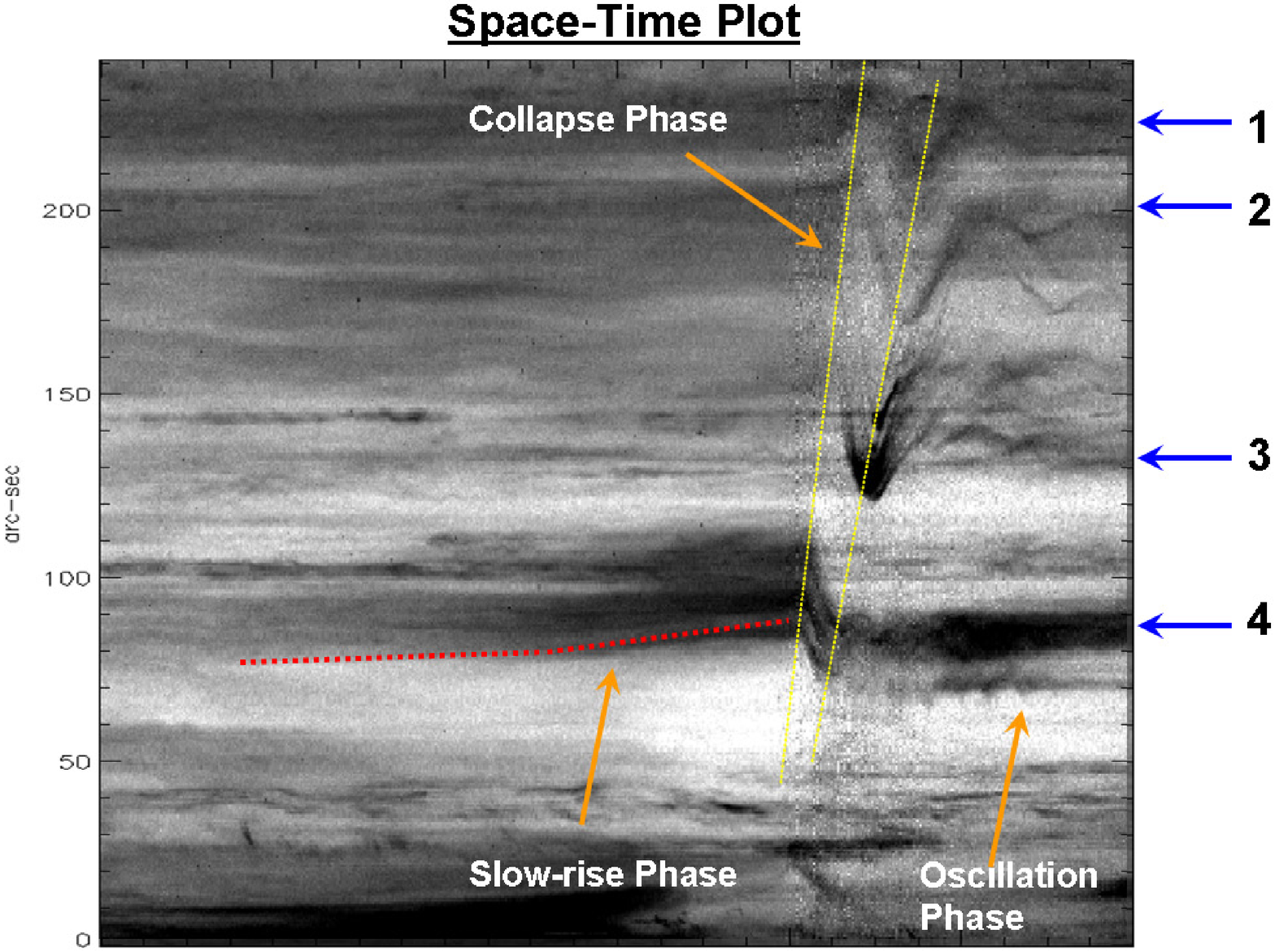}
\epsscale{0.801}
\plotone{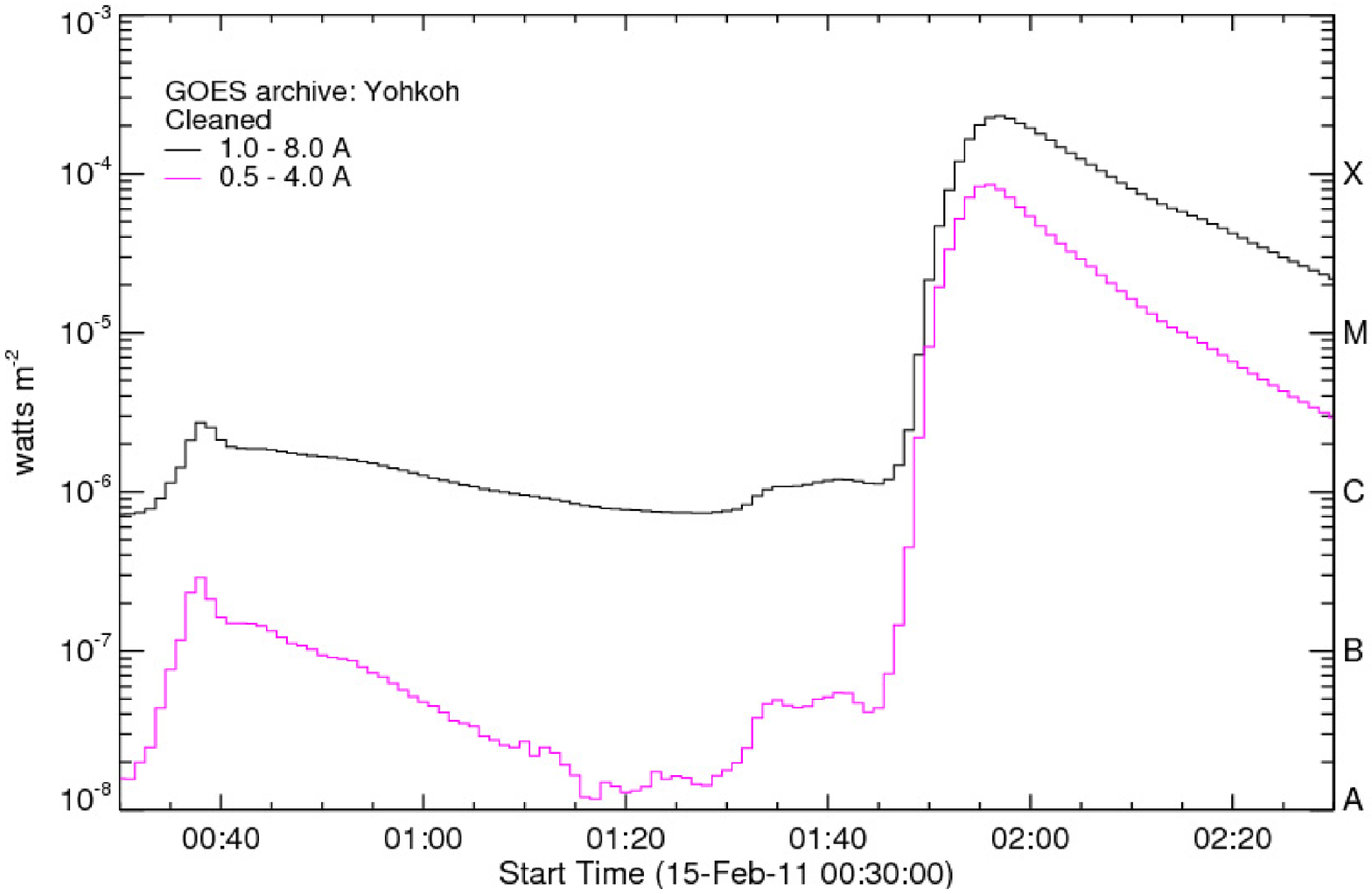}
\caption{The top panel shows the space-time diagram corresponding to the slit marked in Figure 1. The position corresponding to the loops 1-4 are marked on the right side. The three phases of evolution discussed in the text are marked on the figure and indicated by arrows. The bottom panel shows the soft X-ray light curve observed by GOES satellite during the same time interval. \label{fig2}}
\end{figure}

\clearpage

\begin{figure}
\epsscale{0.50}
\plotone{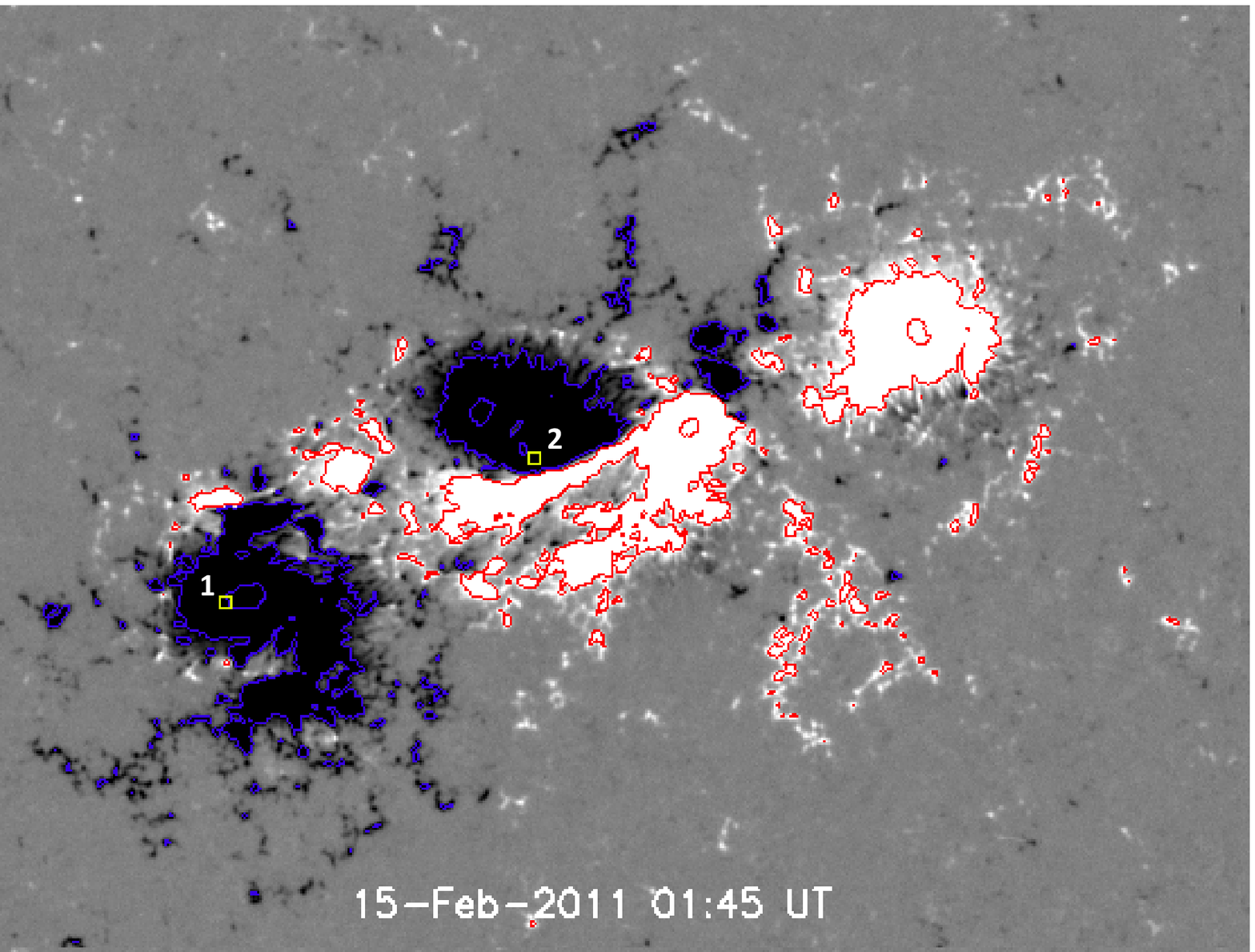}
\plotone{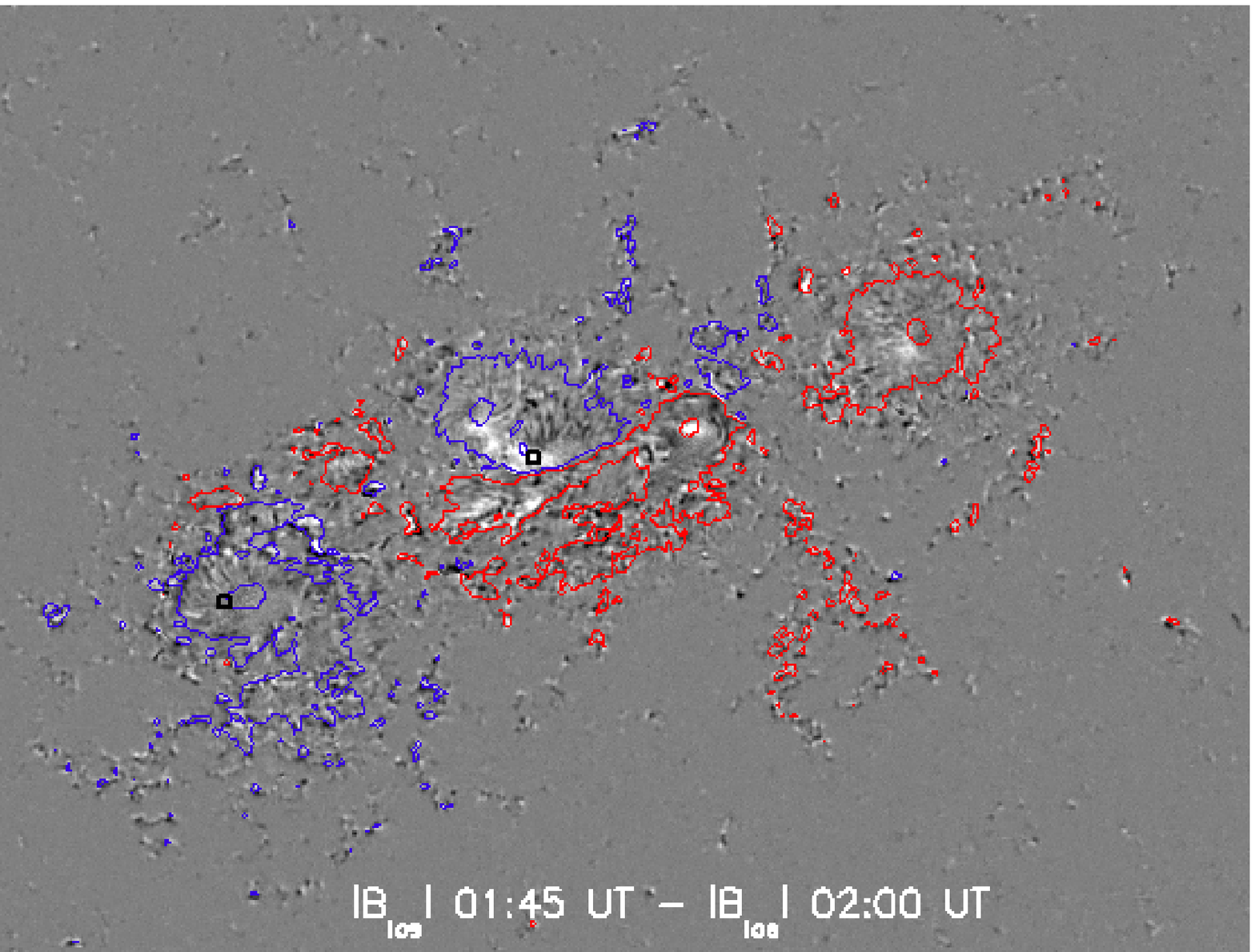}
\caption{The top panel shows the map of longitudinal magnetic field  scaled to  $\pm$ 250 G. The bottom panel shows the absolute difference between the longitudinal magnetograms obtained at 01:45 (pre-flare) and 02:00 UT (post-flare) time interval, scaled between 0 and 250 G. The line contours drawn at 500 and 1000 G levels in both the panels  in blue (red) color represent negative (positive) polarity of the longitudinal field. The boxes `1' and `2' marked in top panel represent the locations of quiet and flaring region, respectively.  The value of the longitudinal field and Stokes profiles observed by HMI averaged over these boxes is examined during the flare interval.   \label{fig3}}
\end{figure}

\clearpage

\begin{figure}
\epsscale{0.70}
\plotone{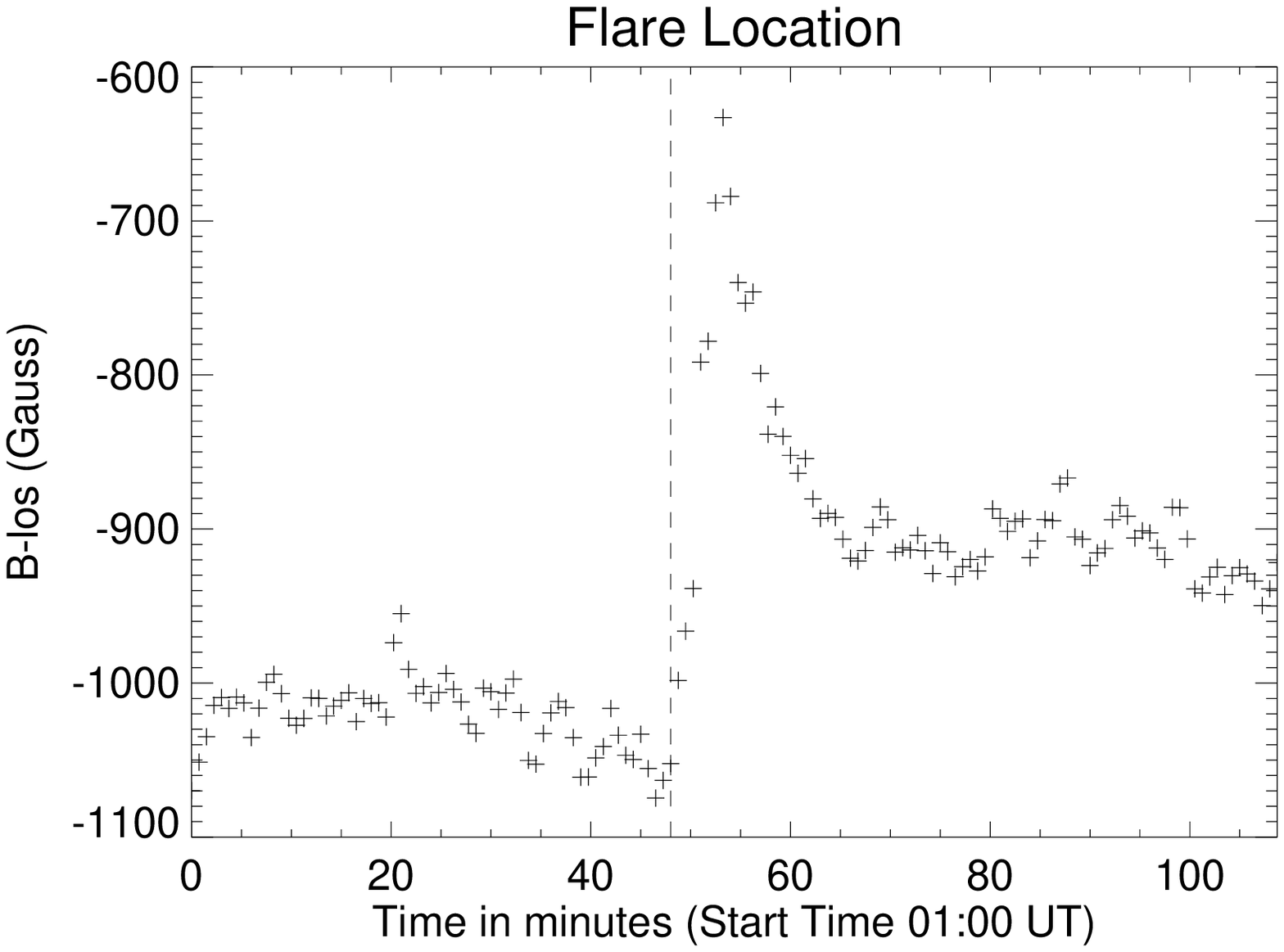}
\plotone{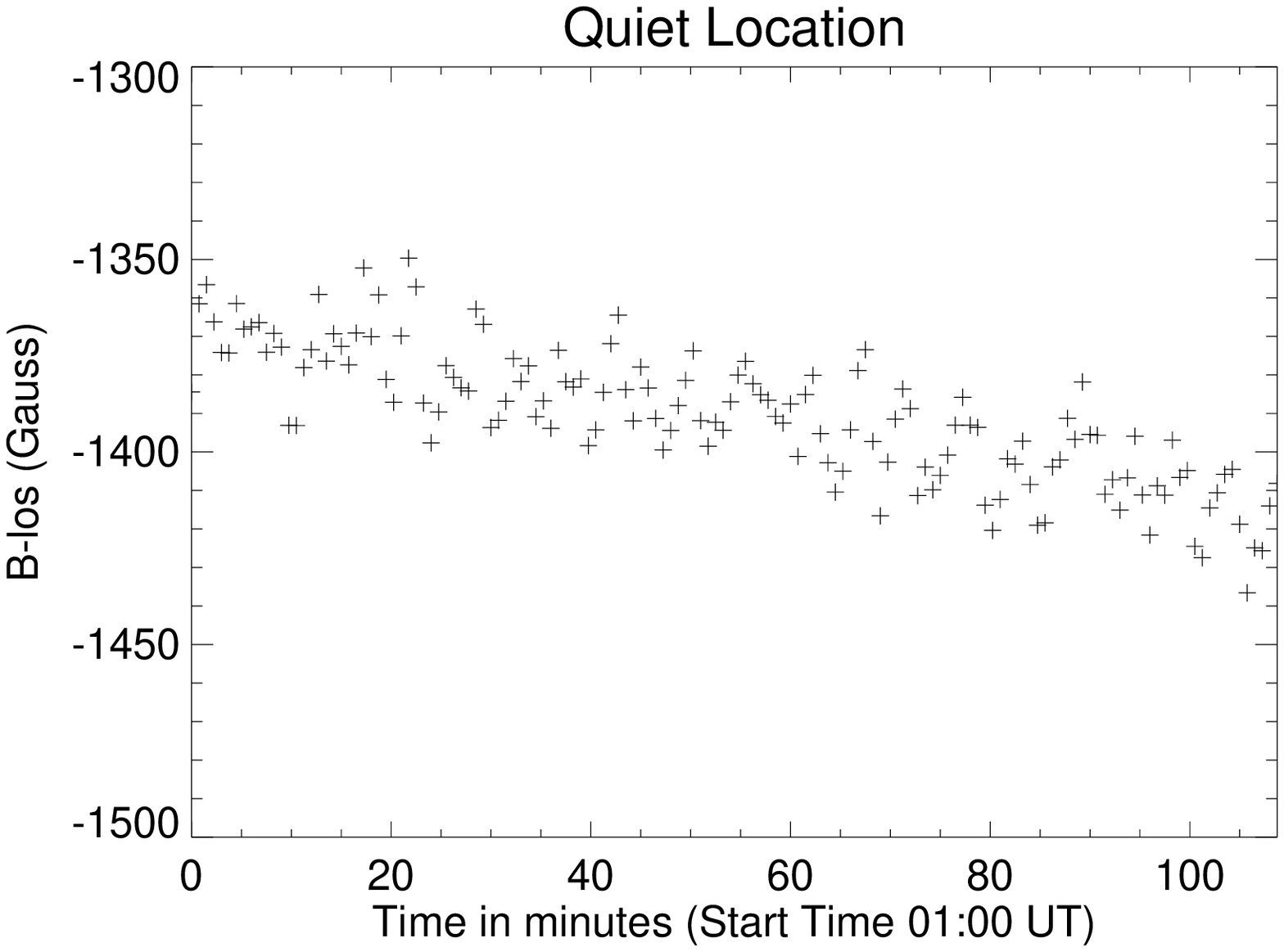}
\caption{The top and bottom panels show the time profile of the average longitudinal magnetic  field inside box `1' and `2' respectively. The box `1' corresponds to the flaring location while box`2' represents a reference quiet region far away from the flaring location. The flaring location in top panel shows abrupt and permanent field change unlike quiet region where the field shows gradual evolution.
\label{fig4}}
\end{figure}

\clearpage

\begin{figure}
\epsscale{0.70}
\plotone{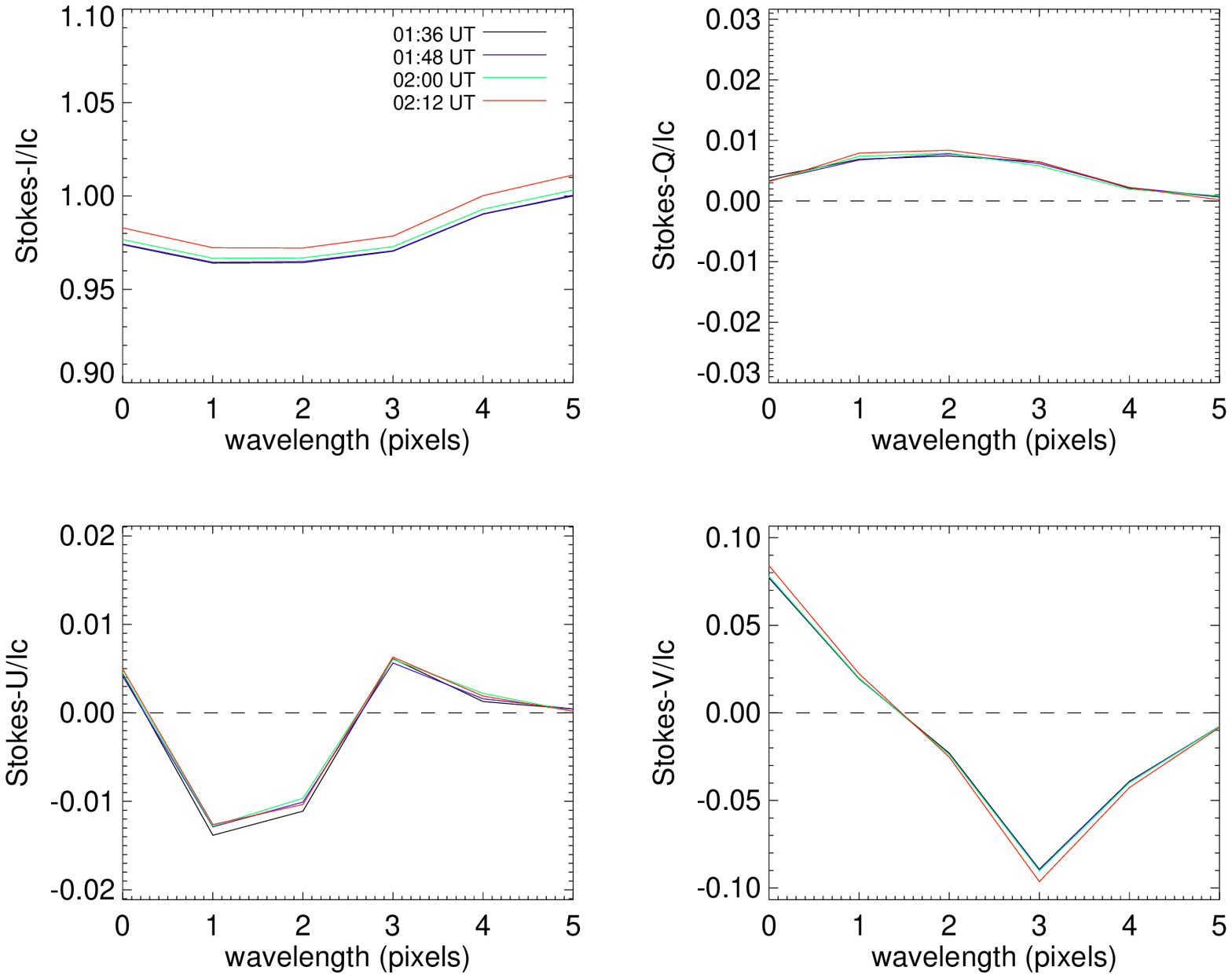}
\plotone{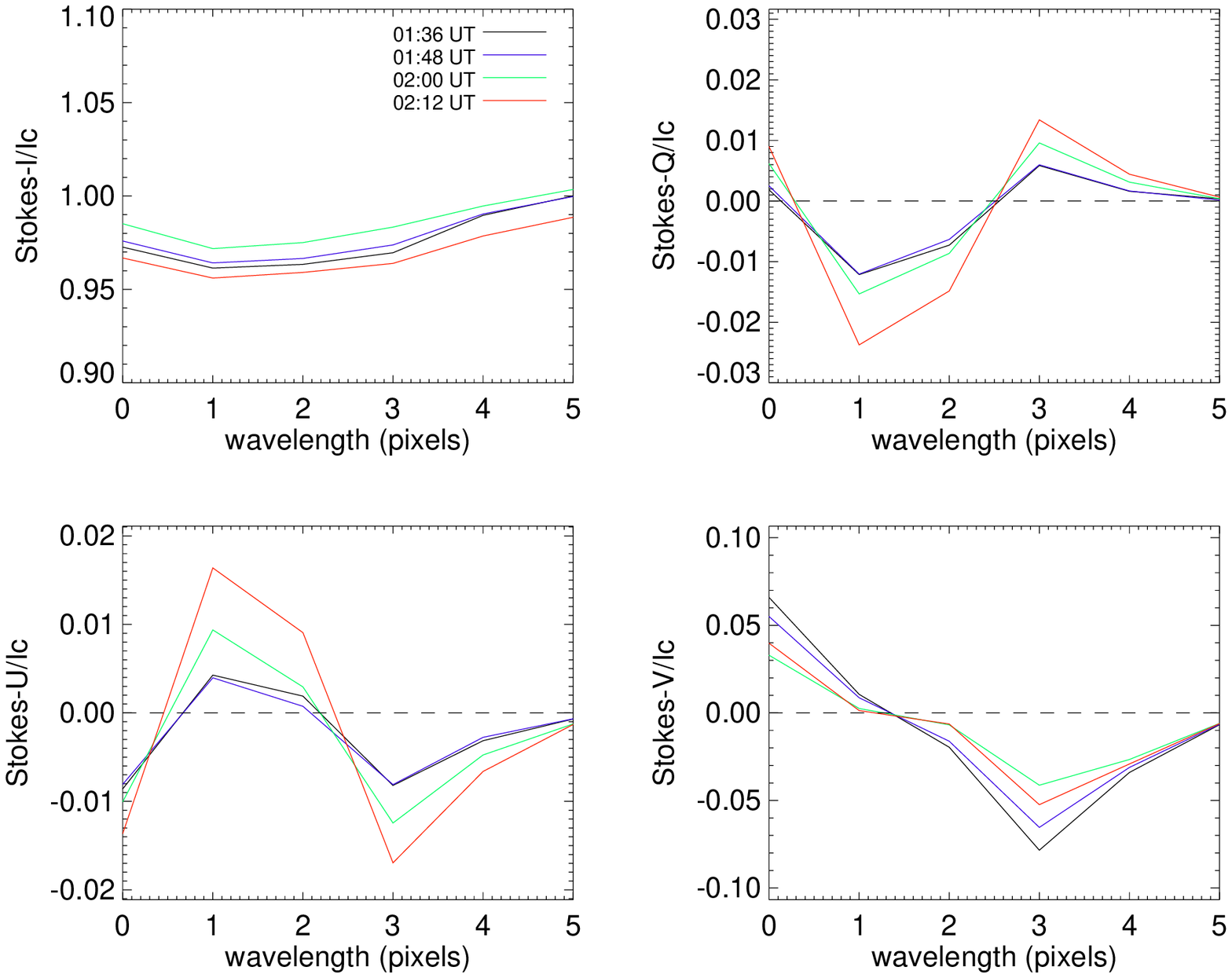}
\caption{The top (bottom) four plots show evolution of the mean Stokes profiles (normalized to continuum intensity, $I_c$) inside the box `1' (box `2') marked in the top panel of Figure 3. The four sets of Stokes $I/I_c$,$Q/I_c$,$U/I_c$ and $V/I_c$ profiles shown in different colors correspond to pre-flare time interval, 01:36, 01:48 (black and blue curves) and post-flare time-interval, 02:00 and 02:12 UT (red and green curves). \label{fig5}}
\end{figure}

\clearpage

\begin{figure}
\epsscale{0.49}
\plotone{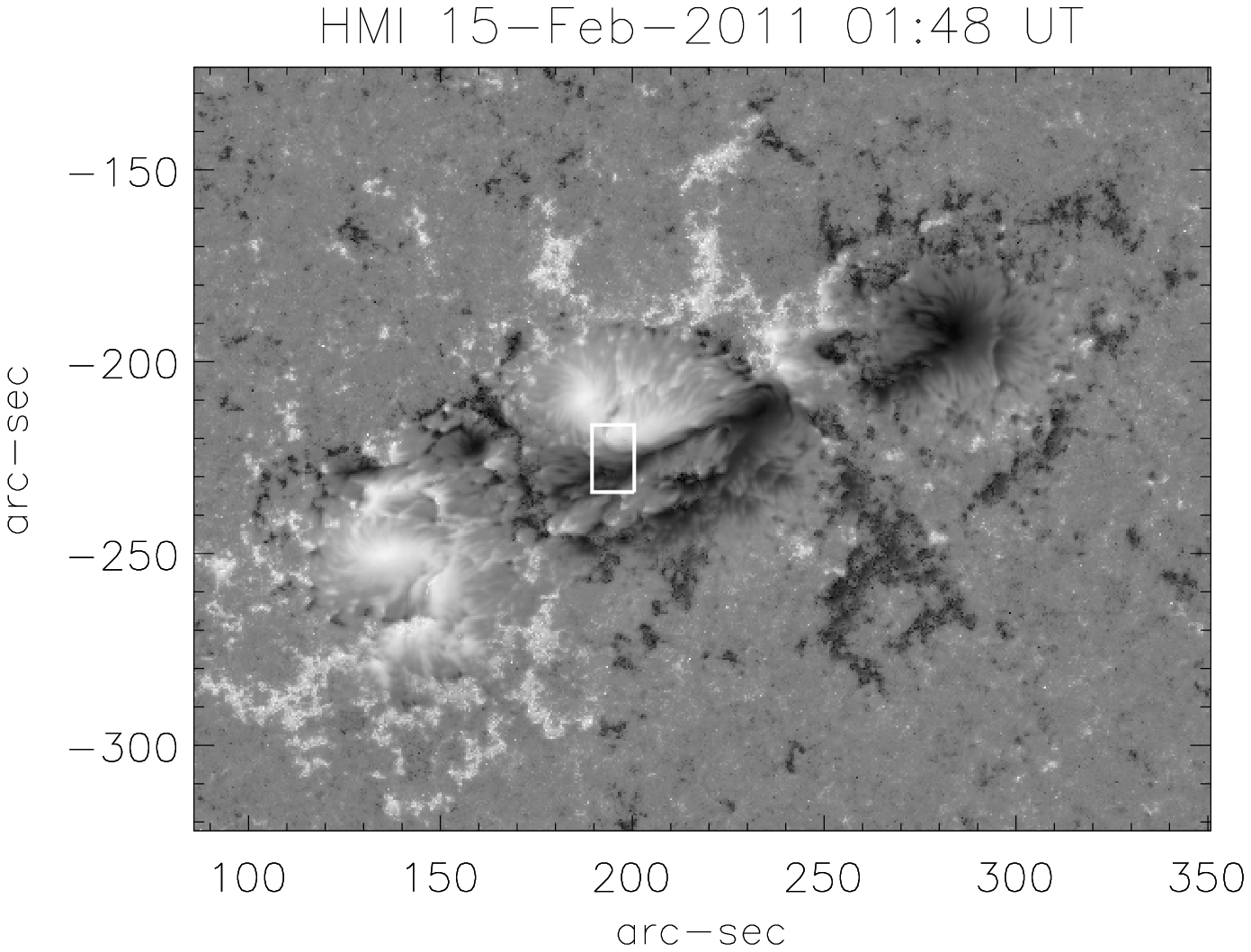}
\plotone{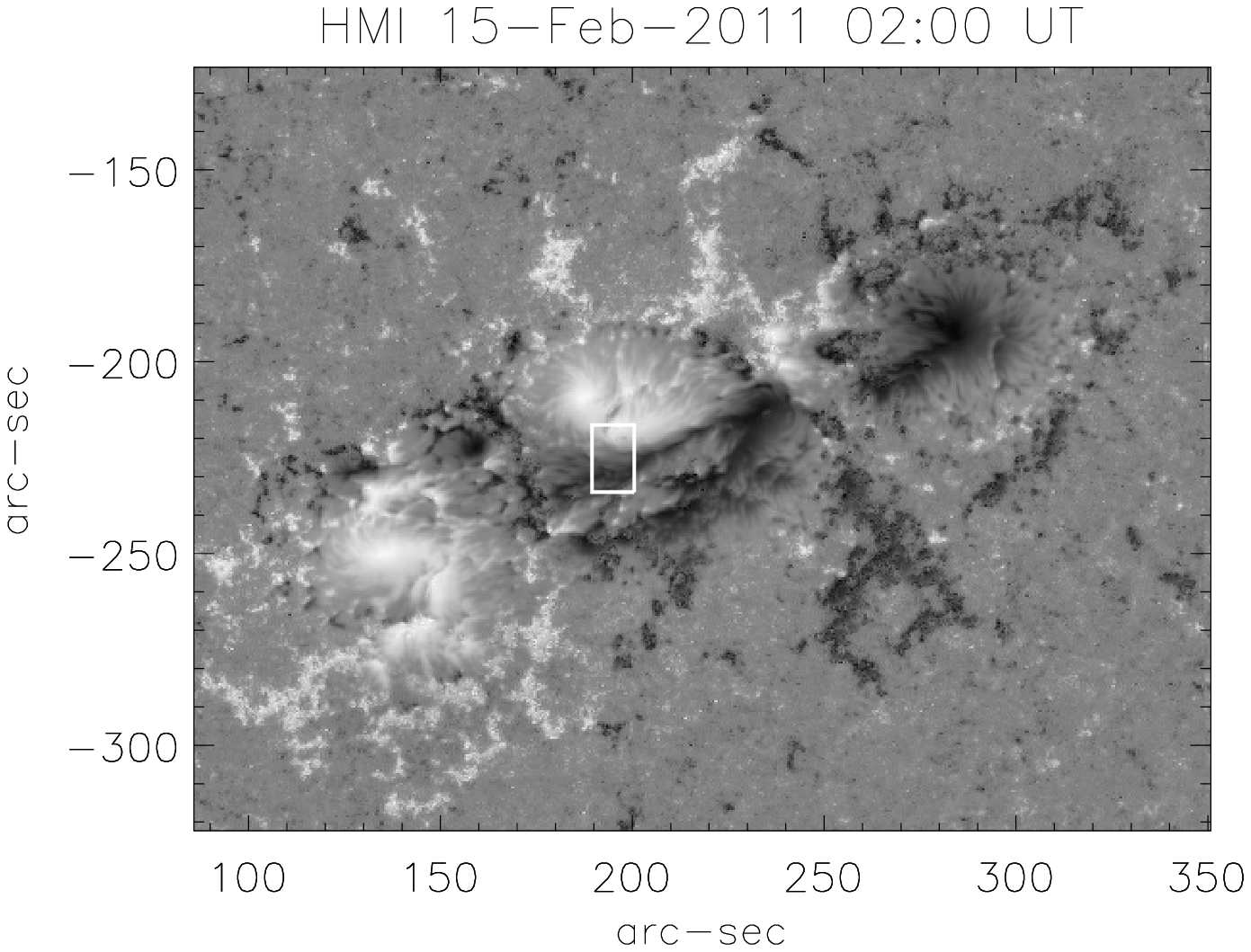}
\plotone{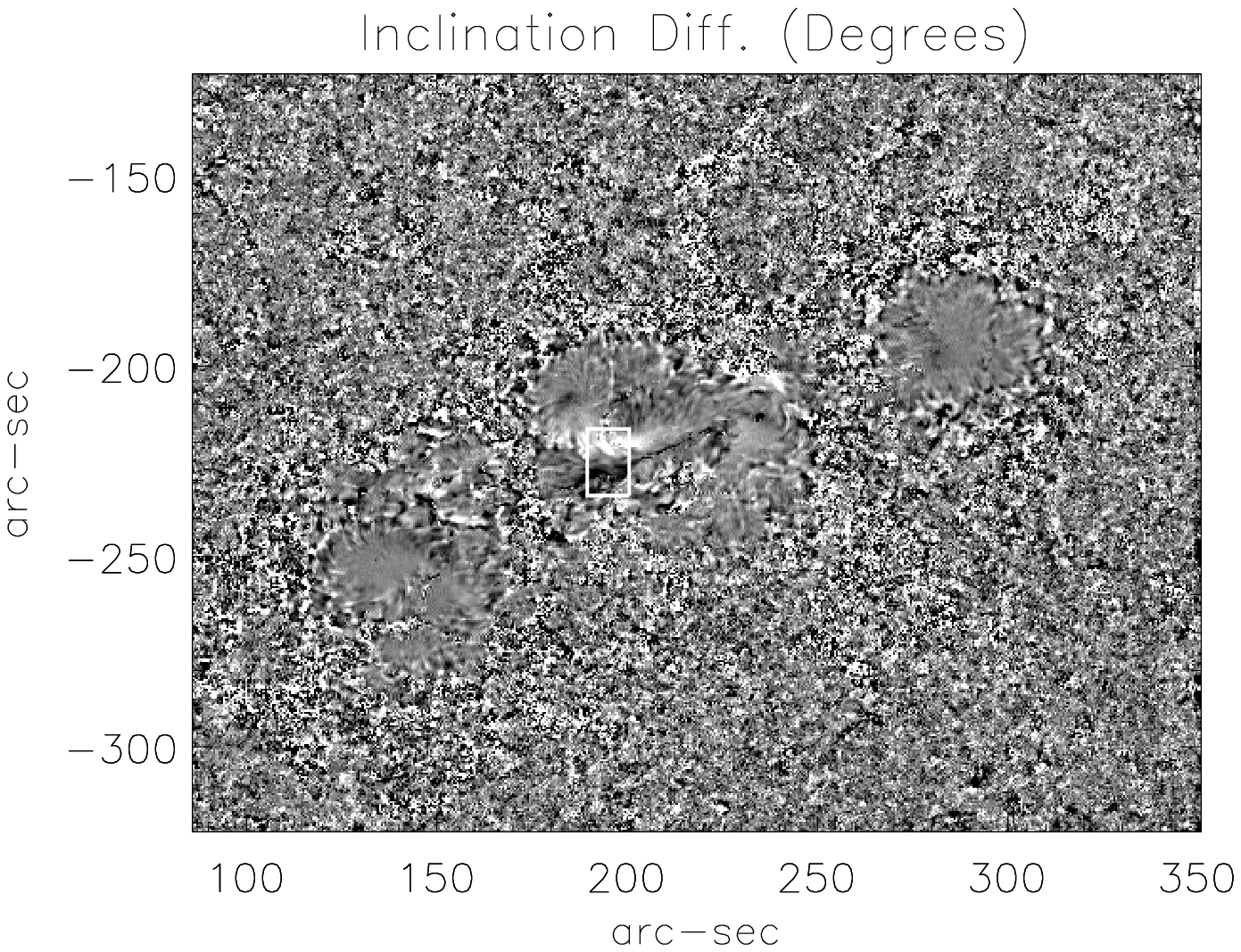}
\plotone{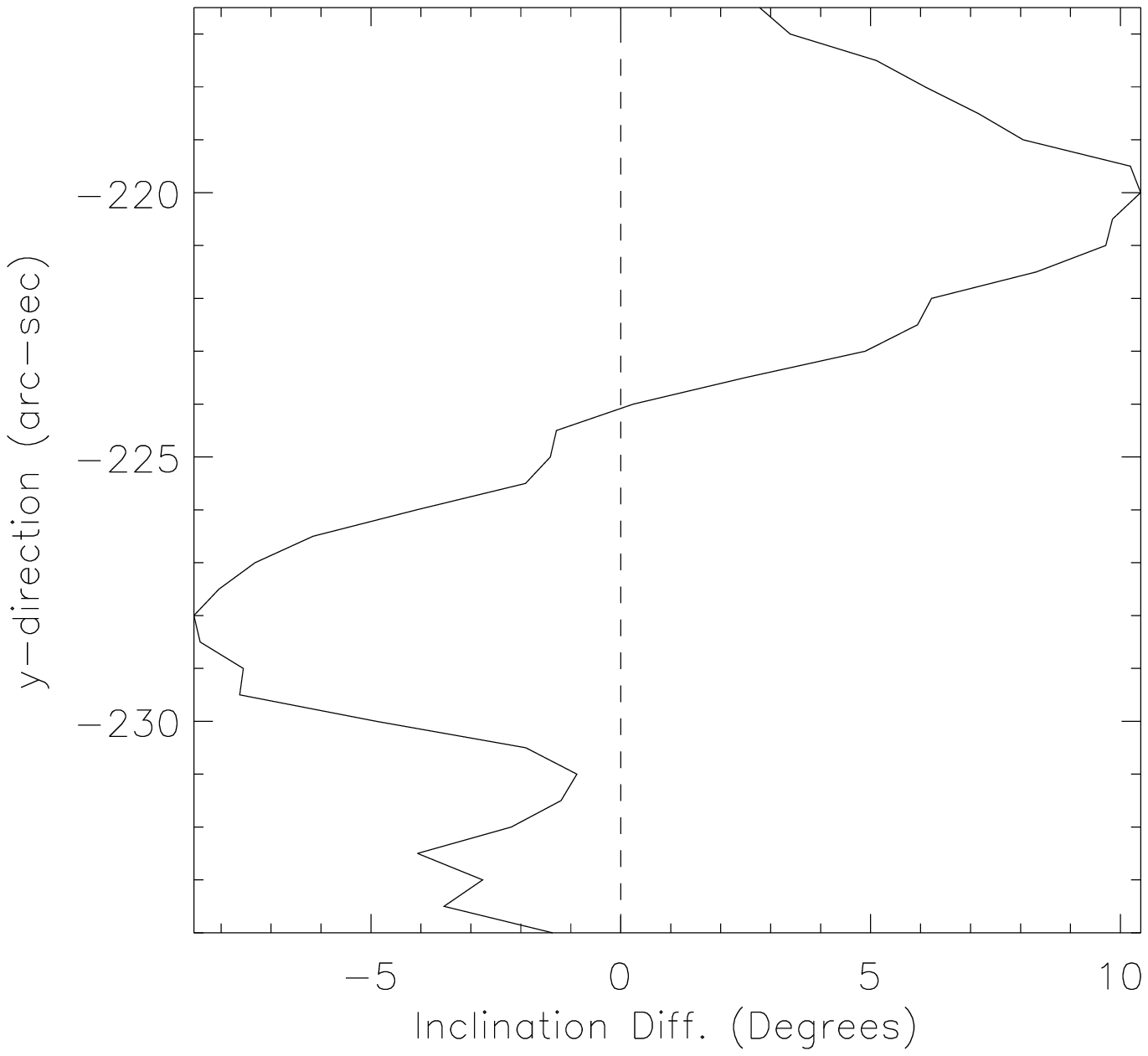}
\caption{The top panels show the maps of magnetic field inclination angle  during times 01:48 and 02:00 UT. The gray levels (black to white) represent inclination angles from 0$^\circ$ (positive polarity) to 180$^\circ$ (negative polarity). The bottom left panel shows the difference (01:48 - 02:00 UT) of the two inclination maps shown in the top panels. The gray levels are scaled between $\pm$10$^\circ$. A rectangular box is marked over the difference map where a change in the inclination angle is observed coherently over a patch near the polarity inversion line (PIL). The mean profile of the change in inclination angle within this box (averaged along abscissa, x-direction) is displayed in lower right panel. \label{fig6}}
\end{figure}

\clearpage


%
%
%

\begin{figure}
\epsscale{1.02}
\plotone{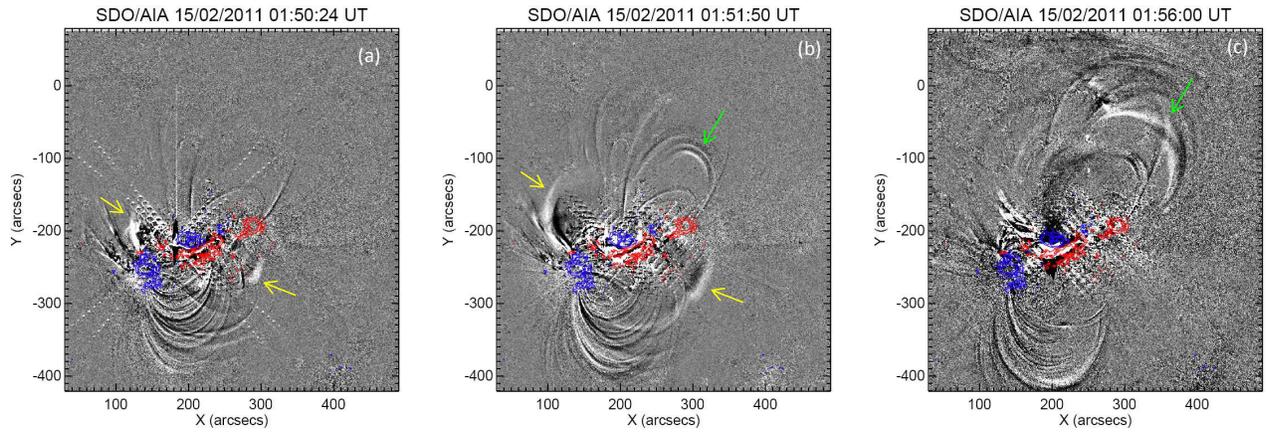}
\caption{The backward running difference of frames, shown at three different times. The first appearance of EUV bright fronts is shown in left panel (yellow arrows).
The middle panel shows the first appearance of collapsing loops (green arrow) and the location of EUV bright fronts (yellow arrows). The right panel shows the collapse of higher loops (green arrow). The line contours at 500 and 1000 G level of the longitudinal magnetic field observed by SDO/HMI instrument are overlaid in blue (red) colors, representing negative (positive) polarity, respectively.  }
\end{figure}

\clearpage

\end{document}